\documentclass[usenatbib,useAMS]{mn2e}
\usepackage{epsf}
\usepackage{epsfig}
\usepackage{amsmath,amssymb}
\usepackage{times}
\usepackage{natbib}
\usepackage{dcolumn}
\usepackage[english]{babel}
\RequirePackage{amsmath}
\RequirePackage{amssymb}        
\RequirePackage{wasysym}

\title[The chemically peculiar hotsubdwarf LS\,IV--14$^{\circ}$116]{An extremely peculiar hot subdwarf with 
 a ten-thousand-fold excess of zirconium, yttrium, and strontium.
}

\author[Naslim, N. et al.]
{
Naslim\,N.$^1$\thanks{E-mail: nas@arm.ac.uk}, 
C.S.\,Jeffery$^1$\thanks{E-mail: csj@arm.ac.uk},  
N.T.\,Behara$^{1,2}$ \&
A.~ Hibbert$^3$\\
$^1$Armagh Observatory, College Hill, Armagh BT61\,9DG\\
$^2$Institut d'Astronomie et d'Astrophysique, Universit\'e Libre de Bruxelles, Belgium\\
$^3$School of Mathematics and Physics, Queens University Belfast,
Belfast BT1 7NN \\
}

\date{Accepted .....
      Received ..... ;
      in original form .....}

\pagerange{\pageref{firstpage}--\pageref{lastpage}}
\pubyear{2010}

\begin{document}

\maketitle

\label{firstpage}

\begin{abstract}
Helium-rich subdwarf B (He-sdB) stars represent a small group of
low-mass hot stars with luminosities greater than those of
conventional subdwarf B stars, and effective temperatures lower than
those of subdwarf O stars. By measuring their surface chemistry,
we aim to explore the connection between He-sdB stars, He-rich sdO stars
and normal sdB stars.  

LS\,IV$-14^{\circ}116$ is a relatively intermediate He-sdB star, 
also known to be a photometric variable. 
High-resolution blue-optical spectroscopy was obtained with the Anglo-Australian
Telescope.
Analysis of the spectrum shows LS\,IV$-14^{\circ}116$  to 
have effective temperature $T_{\rm eff} = 34\,000\pm500 {\rm K}$, 
surface gravity $\log g = 5.6\pm 0.2$, 
and surface helium abundance $n_{\rm He} = 0.16\pm0.03$ by number. 
This places the star slightly above the standard extended hori\-zontal branch, as represented by normal sdB stars. The magnesium and silicon
abundances indicate the star to be metal poor relative to the Sun.

A number of significant but unfamiliar absorption lines were
identified as being due to germanium, strontium, yttrium and zirconium.
After calculating oscillator strengths (for Ge, Y and Zr),
the photospheric abundances of these elements were established to range
from 3 dex (Ge) to  4 dex (Sr, Y and Zr) {\it above} solar. The most
likely explanation is that these overabundances are caused by
radiatively-driven diffusion forming a chemical cloud layer in the
photosphere. It is conjectured that this cloud formation could be
mediated by a strong magnetic field. 
\end{abstract}


\section{Introduction}
Subdwarf B stars are low-mass core helium burning stars with extremely
thin hydrogen envelopes. They behave as helium main-sequence
stars of roughly half a solar mass. Their atmospheres are 
generally helium deficient; radiative levitation and
gravitational settling combine to make helium sink below the
hydrogen-rich surface \citep{heber86}, to deplete other light
elements, and to enhance abundances of heavy elements in the
photosphere \citep{otoole06}.

However, almost 5$\%$ of the total subdwarf population comprise stars
with helium-rich atmospheres \citep{green86,ahmad06}. They have
been variously classified as sdOB, sdOC and sdOD \citep{green86} stars,
but more recently as He-sdB and He-sdO stars
\citep{moehler90,ahmad04}. Their optical spectra are characterised 
by strong He{\sc i} (He-sdB) and He{\sc ii} (He-sdO) lines. 

LS\,IV$-14^{\circ}116$ was classified as an sdO star by
\citet{kilkenny90} and as an He-rich sdO star by
\citet{Viton91}. \citet{jeffery96b} included this star in their
catalogue of He-sdB stars. \citet{ahmad05} reported pulsations in
LS\,IV$-14^{\circ}116$  with periods of 1950 and 2900\,s, the first discovery of
pulsation in a He-sdB star. However, it is only mildly  helium rich,
with $n_{\rm He}\approx0.2$ \citep{ahmad03}.

This paper reports the first detailed abundance analysis of LS\,IV$-14^{\circ}116$
using high resolution optical spectroscopy. In particular, it reports the first discovery
of strontium, germanium, yttrium and zirconium in the optical spectrum of a hot subdwarf.


\section{Observations}

Three spectra of LS\,IV$-14^{\circ}116$ were obtained with the
University College London Echelle Spectrograph (UCLES) 
on the Anglo-Australian Telescope (AAT) on 2005 August 26. Each
exposure was 1800\,s in duration. The observations were obtained, 
reduced and analyzed in the same way as described by \citet{naslim10}. 

\begin{figure}
\includegraphics[angle=-90,width=0.5\textwidth]{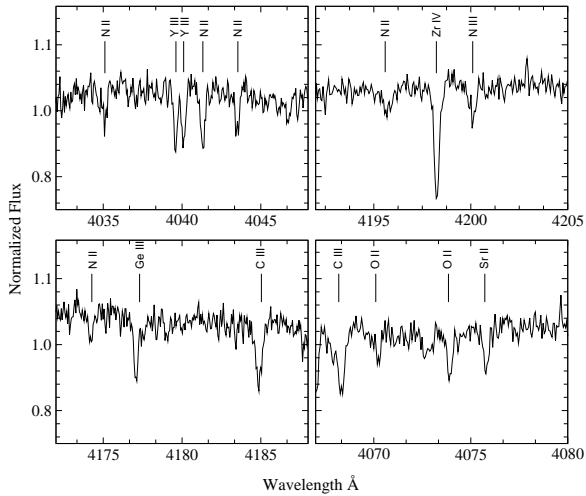}
 \caption{ Y{\sc iii},  Zr{\sc iv}, Sr{\sc ii} and  Ge{\sc iii} lines in LS\,IV$-14^{\circ}116$.} 
  \label{fig_lines}
\end{figure}

The AAT/UCLES spectrum of LS\,IV$-14^{\circ}116$ shows N{\sc ii}, N{\sc
  iii}, C{\sc ii}, C{\sc iii}, O{\sc ii}, Si{\sc iii}, and Mg{\sc ii} 
lines, together with strong He{\sc i} and He{\sc ii}$\lambda$4686.  
Unlike other He-sdB stars, the spectrum displays relatively
strong hydrogen Balmer lines. While analysing this spectrum we
noted a number of strong unidentified lines. Using the NIST database
of atomic spectra, we discovered that these lines are due to
Zr{\sc iv}, Y{\sc iii}, Ge{\sc iii} and Sr{\sc ii} (Fig.~\ref{fig_lines}). As far as we know, 
Zr{\sc iv}, Y{\sc iii}, and Ge{\sc iii} have not been identified in
any other ground-based astronomical spectrum, although Zr and Ge 
have been observed with large overabundances in ultraviolet spectra
of sdB stars \citep{otoole04,otoole06,chayer06,Blanchette08}. The Sr{\sc ii} resonance 
lines are seen in chemically peculiar late-B stars (HgMn stars, for
example), but are completely unknown at late-O or early-B spectral
types. These line identifications immediately pointed to a
remarkable chemical anomaly in this star. However, the anomaly could
not be measured immediately, since no oscillator strengths existed for the optical
Zr{\sc iv}, Y{\sc iii}, and Ge{\sc iii} lines. 

We have not identified lines for any iron-group elements in this
spectrum. There remain a small number of lines which we have been unable to
identify. The most notable have wavelengths 4007.40\AA\ and
4216.20\AA\, with equivalent widths 40 and 34 m\AA, respectively.

\begin{figure}
\includegraphics[angle=-90,width=0.5\textwidth]{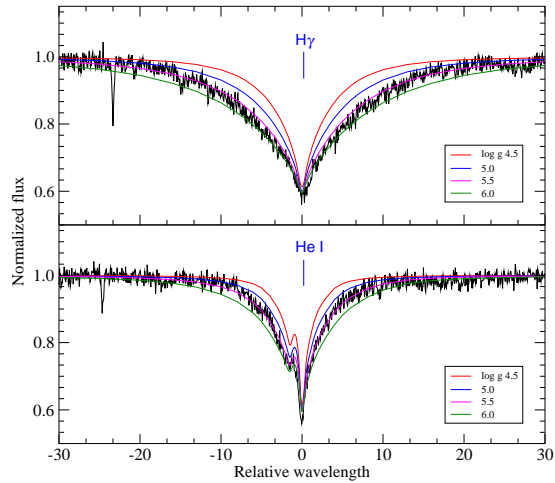}
\caption{Theoretical line profile fits to He{\sc i}\,4471 and
  H$_{\gamma}$ in LS\,IV$-14^{\circ}116$ for model atmospheres with
  $n_{\rm He}=0.100$, $T_{\rm eff}=34\,000\,{\rm K}$ and $\log
  g=4.5(0.5)6.0$.}
  \label{fig_prof}
\end{figure}

\begin{figure}
\includegraphics[angle=-90,width=0.5\textwidth]{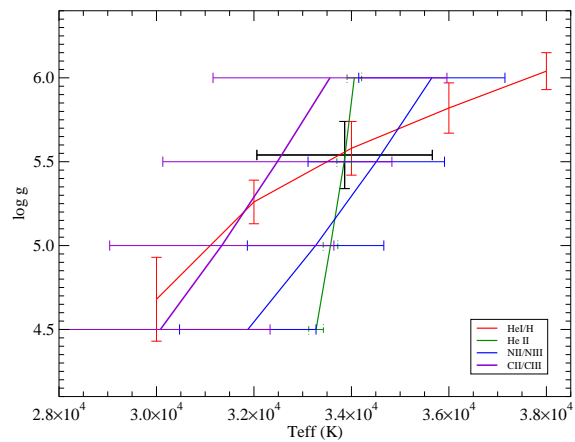}
  \caption{The loci of ionization equilibria for C{\sc ii}/C{\sc iii},
    N{\sc ii}/N{\sc iii} and He{\sc ii}, the profile fits to He{\sc i}
    and H lines, and the adopted solution.}
  \label{fig_ioneq}
\end{figure}

\begin{figure}
\includegraphics[angle=+90,width=0.45\textwidth]{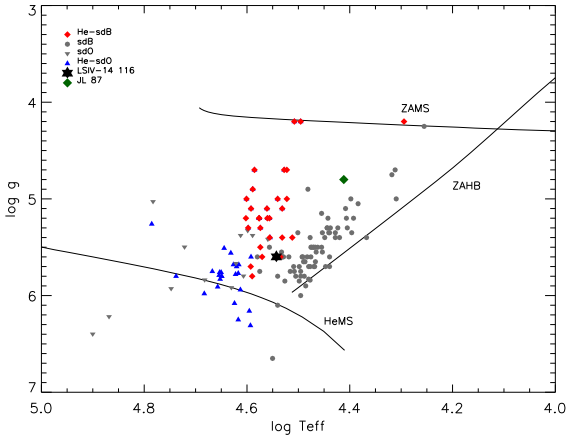}
  \caption{The location of LS\,IV$-14^{\circ}116$ on the $\log g -
    T_{\rm eff}$ diagram, compared with normal sdB stars \citep{edelmann03}
    helium-rich sdB stars \citep{ahmad03,naslim10}, helium-rich sdO
    stars \citep{stroeer07}, and JL\,87 \citep{ahmad07}. A zero-age
    main sequence ($Y=0.28,Z=0.02$), a zero-age horizontal branch
    ($M_{\rm c}=0.485 {\rm M_{\odot}}, Y_{\rm e}=0.28, Z=0.02$) and a
    helium main sequence ($Z=0.02$) are also shown.} 
  \label{fig_gt}  
\end{figure}

\section{Physical parameters of LS\,IV$-14^{\circ}116$}

We measured the effective temperature $T_{\rm eff}$, surface gravity $\log
g$, and elemental abundances of LS\,IV$-14^{\circ}116$
using the same method as described by \citet{naslim10}. 
We found the microturbulent velocity $v_{\rm t} = 10\,{\rm
  km\,s^{-1}}$ by minimising the dispersion on carbon abundance
measured from six C{\sc iii} absorption lines; adopting 
$v_{\rm t} = 5\,{\rm km\,s^{-1}}$ increases this dispersion by 25\%.
The grids of model atmospheres adopted for analysis were computed 
with 1/10 solar metallicity, relative helium abundances $n_{\rm He} =
0.10$, 0.20 and 0.50 by number, and microturbulent velocity  
$v_{\rm t} = 10\,{\rm km\,s^{-1}}$.

$T_{\rm eff}$ and $\log g$ were measured using ionisation
equilibrium and line profile fitting. 
Using appropriate model atmospheres, the ionisation
equilibrium was established by balancing the abundances determined
from N{\sc ii} and N{\sc iii} and from C{\sc ii} and C{\sc iii} 
lines in the spectrum, as well as by
fitting the equivalent width of the temperature sensitive He{\sc ii}\,4686
line.
The surface gravity was established by fitting theoretical
profiles to H$_{\beta}, \gamma$, and $\delta$, and to He{\sc i}\,4471  
 (Fig.~\ref{fig_prof}). 
The coincidence of profile fits and ionisation
equilibria was used to determine the overall solution illustrated in
Fig.~\ref{fig_ioneq}.

\begin{table}
\centering
\caption{Atmospheric parameters}
\label{t_pars}
\begin{tabular}{@{}llll}
\hline
$T_{\rm eff} (\rm K)$& $\log g$      & $n_{\rm He}$ & Source\\
\hline
 $34\,000\pm500    $& $5.6\pm0.1$   & $0.160\pm0.03$ & SFIT \\
 $33\,860\pm1\,800 $& $5.54\pm0.2$ &       & Ion eq \\
 $33\,000\pm1\,000 $& $5.8\pm0.2$   & 0.2   & \citet{Viton91} \\
 $32\,500\pm150    $& $5.4\pm0.1$   & 0.21  & \citet{ahmad03} \\
\hline
\end{tabular}
\end{table}

We also measured the physical parameters $T_{\rm eff}$, $\log g$ and
helium abundance $n_{\rm He}$ using the package SFIT which finds the
best-fit solution within a grid of synthetic spectra.  The model grid
was defined with $T_{\rm eff}=32\,000$ $(2000)$ $38\,000\,{\rm K}$,
$\log g = 4.50 (0.5) 6.00$, $n_{\rm He}=0.10, 0.20, 0.30, 
0.50$ and $v_{\rm t}=10\,{\rm km\,s^{-1}}$.
  The observed spectrum of LS\,IV$-14^{\circ}116$ is
shown along with the best fit model spectrum in Fig.~\ref{fig_AATfit}. 

The atmospheric properties of LS\,IV$-14^{\circ}116$ deduced from
these measurements are given in Table ~\ref{t_pars}. 
The projected rotational line broadening measured from Zr{\sc iv}
  lines 4198.26\AA\, and 4317.08\AA\, is $v \sin i < 2\,{\rm
    km\,s^{-1}}$.
SFIT and ionization equilibrium results agree to
within the formal errors. Although all ionisation equilibria and
profile fits do not converge on a single point, the 
intersections lie within 3200\,K in $T_{\rm eff}$ and
0.5 in $\log g$. We took a weighted mean of these intersections 
to determine the ionization $T_{\rm eff}$ and $\log g$.
For subsequent abundance measurements, we adopted the grid model atmosphere 
with $T_{\rm eff}=34\,000\, {\rm K}$, $\log g=5.5$, $n_{\rm He}=0.20$ 
and $v_{\rm t}=10\,{\rm km\,s^{-1}}$. 

In view of the atmospheric abundances reported below, the above
procedure was repeated with a grid of models in which the abundances of
all metals with atomic number $\geq 20$ (calcium) except iron were
increased to 10 times the solar value. This has proved to be a
successful strategy for the analysis of normal sdB star spectra
\citep{otoole06,pereira10}. However in this case, 
a) we failed to find any agreement between He{\sc II}4686 and the N{\sc ii/iii} ionisation 
equilibrium, b) we found the depth of the theoretical H and He{\sc
  i} lines to be consistently deeper than observed and
c) whilst leading to a 2000\,K reduction in $T_{\rm eff}$, the new
models also led to a significant increase in the measured strontium abundance. 
Without further diagnostics
of high-Z abundances and chemical stratification, the 1/10 solar
metallicity models were deemed satisfactory for the present paper.


\begin{table}
\centering
\caption{Individual line equivalent widths $W_{\lambda}$ and abundances $\epsilon_i$, with
  adopted oscillator strengths $gf$, for LS\,IV$-14^{\circ}116$.}
\label{t_lines}
\begin{tabular}{@{}ccrlc}
\hline
Ion & & & \\
$\lambda/{\rm \AA}$ &
$\log gf$ &&
$W_{\lambda}/{\rm m\AA}$ &
$\epsilon_{i}$  \\
\hline
  C{\sc ii}      &           &&       &           \\
4074.48 & 0.204&$\rceil$  &  42      & 8.20    \\
4074.52 & 0.408  &          &          \\
4074.85 & 0.593&$\rfloor$  &          &            \\
4075.85 & 0.756&$\rceil$  &  57      & 8.08           \\
4075.94 &$-0.076$&$\rfloor$&          &              \\
4267.02 & 0.559&$\rceil$  &  130     & 7.80     \\
4267.27 & 0.734&$\rfloor$  &          &          \\

  C{\sc iii}      &           &&       &         \\
4067.94 & 0.827  &&  60     & 7.70      \\
4068.91 & 0.945  &&  59     & 7.69        \\
4070.26 & 1.037  &&  82     & 8.37       \\
4186.90 & 0.924  &&  76     & 8.34      \\
4647.42 & 0.072  &&  130  &8.08         \\
4650.25 &$-0.149$&&  98   &  8.01       \\
4651.47 &$-0.625$&&  54     & 7.99        \\

 N{\sc ii}       &           &&       &         \\
3995.00 & 0.225  &&  47  &   7.65      \\
4041.31 & 0.830  &&  45  &   7.82     \\
4043.53 & 0.714  &&  46   &  7.95       \\
4236.86 & 0.396&$\rceil$  &  73   &  8.19           \\
4236.98 & 0.567&$\rfloor$  &    &         \\
4241.78 & 0.728&$\rceil$  &  35  &  7.85     \\
4241.79 & 0.710&$\rfloor$  &    &           \\
4447.03 & 0.238  &&  48   & 8.04       \\
4601.48 &$-0.385$&&   35   & 8.15       \\
4607.16 &$-0.483$&&   42   &  8.35      \\
4630.54 &  0.093 &&  50   & 7.88        \\
  N{\sc iii}      &            &&        &        \\
4640.64 & 0.140  &&  77   & 8.19        \\
4634.14 &$-0.108$&&  51   &  8.13      \\

 OII      &            &&        &         \\
4072.15 & 0.545  &&  21          &  7.42           \\   
4414.90 & 0.210  &&  33          &  7.63           \\   
4416.97 &$-0.041$&&  26          &  7.75           \\ 
  
 Mg{\sc ii}      &      &&        &                \\
4481.13 & 0.568 &&  25    &6.85    \\

 Si{\sc iii}      &            &&        &         \\   
4552.62 & 0.283 &&  29   & 6.32   \\
\hline
\end{tabular}\\
\parbox{70mm}{
$gf$ values: 
C{\sc ii} \citet{yan87}, 
C{\sc iii} \citet{hib76,har70,boc55}, 
N{\sc ii} \citet{bec89}, 
N{\sc iii} \citet{but84}, 
O{\sc ii} \citet{Bec88},
Si{\sc iii} \citet{bec90}, 
Mg{\sc ii} \citet{wie66}} \\
\end{table}

\begin{table*}
\centering
\caption{Atomic data for optical lines of Sr, Ge, Y and Zr, and
  abundances derived for LS\,IV$-14^{\circ}116$.}
\label{t_atomic}
\begin{tabular}{@{}ccccrlccc}
\hline
Ion &  &  & & & & \\
$\lambda/{\rm \AA}$ & Configuration & $E_{i}/{\rm eV}$ & Ref & $\log
gf$ && Ref & 
$w_{\lambda}/{\rm m\AA}$ & $\epsilon_{i}$  \\
\hline
Sr{\sc ii}      &      &     &&         &      &            \\
4077.71  &$5s^2S_{1/2} - 5p^2P_{3/2}$    &\,0.00&1&  0.142 &&1& 24 & 7.00 \\
4215.52  &$5s^2S_{1/2} - 5p^2P_{1/2}$    &\,0.00& &$-0.175$&& & 10 & 6.91 \\[1mm]
Ge{\sc iii}      &      &     &           &&      &     &       \\
4178.96 &$4s5s ^3S_1  -  4s5p ^3P_2$    & 19.66&2&  0.341 &&5& 63 & 6.36 \\
4260.85 &$4s5s ^3S_1  -  4s5p ^3P_1$    & 19.66& &  0.108 && & 40 & 6.34 \\
4291.71 &$4s5s ^3S_1  -  4s5p ^3P_0$    & 19.66& &$-0.368$&& & 10 & 6.14 \\[1mm]
 Y{\sc iii}       &     &       &               &&        & &            \\
4039.602&$4f^2F_{7/2} - 5g^2G_{9/2}$     & 12.53&3& 1.005&$\rceil$ &5& 21 & 6.12  \\
4039.602&$4f^2F_{7/2} - 5g^2G_{7/2}$     & 12.53& &$-0.538$&$\rfloor$& &  & \\
4040.112&$4f^2F_{5/2} - 5g^2G_{7/2}$     & 12.53& &  0.892 && & 21 & 6.24 \\[1mm]
Zr{\sc iv}        &     &       &               &&        &     &       \\
4137.435&$5d^{2}D_{3/2}-6p^{2}P_{3/2}^{0}$& 18.18&4&$-0.625$&&5& 10 & 6.20 \\
4198.265&$5d^{2}D_{5/2}-6p^{2}P_{3/2}^{0}$& 18.23& &  0.323 && & 86 & 6.43 \\
4317.081&$5d^{2}D_{3/2}-6p^{2}P_{1/2}^{0}$& 18.18& &  0.069 && & 64 & 6.49 \\
4569.247&$5g^{2}G_{7/2}-6h^{2}H_{9/2}^{0}$& 25.65& &  1.127&$\rceil$ & &101 & 6.76 \\
4569.247&$5g^{2}G_{9/2}-6h^{2}H_{11/2}^{0}$&25.65& &  1.216&$\rfloor$& &    &  \\
\hline
\end{tabular}\\
\parbox{100mm}{
References:1. \citet{brage98}, 
2. \citet{lang29},
3. \citet{epstein75},
4. \citet{reader97},
5. this paper. }
\end{table*}

\section{Atomic Data}

In order to measure elemental abundances for all visible species, 
we measured equivalent widths of lines of C{\sc ii},C{\sc iii}, 
N{\sc ii}, N{\sc iii}, O{\sc ii}, Mg{\sc ii}, Si{\sc iii}, Sr{\sc ii},
Ge{\sc iii}, Y{\sc iii} and Zr{\sc iii}. Atomic data and oscillator
strengths for most of these lines were available in our own
compilation (see Table \ref{t_lines}), but data for  
Sr, Ge, Y and Zr were not. 

The atomic data for the relevant transitions of these ions are 
displayed in Table \ref{t_atomic}. The labels of the upper and 
lower states of each transition include,
in addition to the angular momenta and parity, the electron occupancy
of the valence subshells, as given by the Hartree-Fock approximation.
In practice, the Hartree-Fock (HF) approximation is not sufficiently
adequate for the calculation of oscillator strengths, as demonstrated
by the work of \citet{brage98}
for Sr{\sc ii}.  Instead, each wave
function needs to be represented by a configuration interaction (CI)
expansion, though the dominant (largest CI mixing coefficient) remains
the HF configuration in all the cases considered here, and therefore
constitutes an appropriate part of the label. 

For the Sr{\sc ii} transitions, we have merely quoted the oscillator
strength results of \citet{brage98}, since their work is a
substantial MCHF calculation and their final results agree well with
the experimental studies of \citet{grevesse91}
and even with the conceptually simpler calculation of \citet{migdalek79}
who represented core polarisation by model potentials rather than by
explicit CI.  However, transition data for the other ions has not been
calculated to such accuracy, and therefore we undertook to provide
that data ourselves. 

For ions with valence electrons outside a number of closed subshells,
there are three types of electron correlation which improve upon the
HF approximation: valence shell correlation (with the core described
as in the HF approximation); core-valence correlation (or core
polarisation), in which the influence of the core on the valence
electrons is included; core-core correlation.  For monovalent ions,
only the second and third of these effects occurs, in accordance with
the work of \citet{brage98}.  We describe below our work on
the three ions Ge{\sc iii}, Y{\sc iii}, Zr{\sc iv}.

\subsection{Ge {\sc iii}}

This is the only ion considered here which has two valence electrons
and therefore the only ion in which valence shell correlation must be
included.  Much work has been published on transitions amongst the
$n=4$ levels \citep[for example,][]{chen10}, 
but almost nothing on transitions involving higher levels. We
undertook our calculations of oscillator strengths using the
configuration interaction code CIV3 \citep{hibbert75,hibbert91}.

Relativistic effects, giving rise to fine-structure splitting, were
included using the Breit-Pauli approximation, but all orbital
optimisations were carried out in $LS$ coupling.  The radial functions
of the orbitals are expressed in analytic form as sums of Slater-type
orbitals.  We used (in the absence of functions for Ge{\sc iii}) the radial
functions of the ground state of Ge{\sc ii} given by \citet{clementi74}
but reoptimised the outer orbital functions $4s$ and $4p$ on the
energy of the $4s4p$£ $^3$P$^o$ state of Ge{\sc iii}.  With these orbitals
fixed, we optimised $5s$ and $5p$ on the energies of $4s5s$ $^3$S and
$4s5p$ $^3$P$^o$.  Subsequent to this, we optimised $4d$ and $4f$ on
the energy of $4s5p$ $^3$P$^o$ and $5d$ on $4s5s$ $^3$S.
Valence-shell correlation was represented by constructing all possible
configurations using these orbitals, but keeping a common core of
filled subshells up to $3d$. 

Core-valence correlation was represented by configurations with a
single electron removed from $3d^{10}$ and replaced by $6p$ or $5f$,
optimised on $4s5s$ $^3$S, with $4s5s$ replaced by $4s4p/4s5p$ and
vice versa.  Core-core correlation is likely to have a very small
effect on the oscillator strengths. 

Our oscillator strengths (length form) are shown in Table \ref{t_atomic}, and are
obtained using experimental wavelengths.  The calculated energies
agree with experiment to within 2\%.  The length and velocity forms
also agree to within 5\%.  We also undertook the corresponding
calculations in $LS$ coupling and found approximately the same level
of accuracy.  We noted that, unlike the Sr{\sc ii} work, core polarisation
had only a small effect. This is partly because the orbitals of the
transition ($5s$ and $5p$) are further from the core, and partly
because the effective core experienced by these electrons also
includes the $4s$ electron. 

\subsection{Y {\sc iii}} 

We proceeded in a similar manner for Y{\sc iii}: the starting point were
the radial functions for the ground state of Y{\sc ii} given by Clementi
and Roetti (1974), and then reoptimised $4s$, $4p$ along with $4f$ on
the $4s^24p^64f$ $^2$F$^o$ state, and then the $5g$ orbital function
on the $^2$G state.  In this case, no valence-shell correlation is
possible, so core polarisation is the dominant correlation effect
modelled explicitly by opening the $4p$ and $4s$ subshells.
Specifically, we added to $4s^24p^65g$ $^2$G the configurations
$4s^24p^5(4d6h+4d4f+4d5f)$ and $4s4p^6(4f5p+5p5f+5p6h)$, with similar
additions for the $^2$F$^o$ state.  We also added limited core-core
correlation by means of configurations in which $4s$ is replaced by
$4d$ in the main configurations, as well as $4p^6$ by $4p^44d^2$.

The resulting oscillator strengths are shown in Table \ref{t_atomic}. Again,
calculated transition energies agree to with 4\% of experiment.  We
consider that the oscillator strengths are accurate to better than
5\%.  As with Ge{\sc iii}, the effect of core polarisation is quite small,
which is not surprising in view of the higher orbital angular momenta
of the outer electron. 

\subsection{Zr {\sc iv}}

A similar process to that adopted for Y{\sc iii} was carried out for
Zr{\sc iv}, with the core orbitals chosen as those of the Zr{\sc ii} 
ground state, 
while $4s$, $4p$ were reoptimised along with $5g$ on the Zr{\sc iv}
configuration $5g$ $^2G$ and, with these core orbitals, $6h$ was
optimised on the $6h$ $^2$H$^o$ state.  In a similar manner to Y{\sc iii},
orbitals $4d,~5s,~5p$ were introduced to allow for core
polarisation. Two other orbitals, $4f$ and $7i$, were introduced to
allow for a more angular flexibility in the outer electrons.  Again
the effects of core polarisation are small, and length and velocity
forms of the oscillator strengths agreed to around 3\%. 

A separate
calculation was undertaken for the $5d$--$6p$ transitions.  We
reoptimised $4p$ along with $4d$ on the $^2$D state, and then with
these functions optimised $5d$, $5p$ and $6p$ on their respective
states. To represent core-valence correlation, additional orbitals
$5s,~7p,~6d,~4f$ were optimised to allow for configurations with one
electron removed from the $4p$ subshell and $8p$ in order to polarise
the $4s$ subshell.  The calculated transition energies agreed very
well with experiment (to within 1\%).  We estimate that the oscillator
strengths are correct to within better than 10\%.  The effect of core
polarisation was still small, principally due to the high $n$-values
of the outer electrons.

\section{Abundances}

\begin{figure*}
\centering
\includegraphics[trim=0cm 3cm 0cm 0cm,width=18cm]{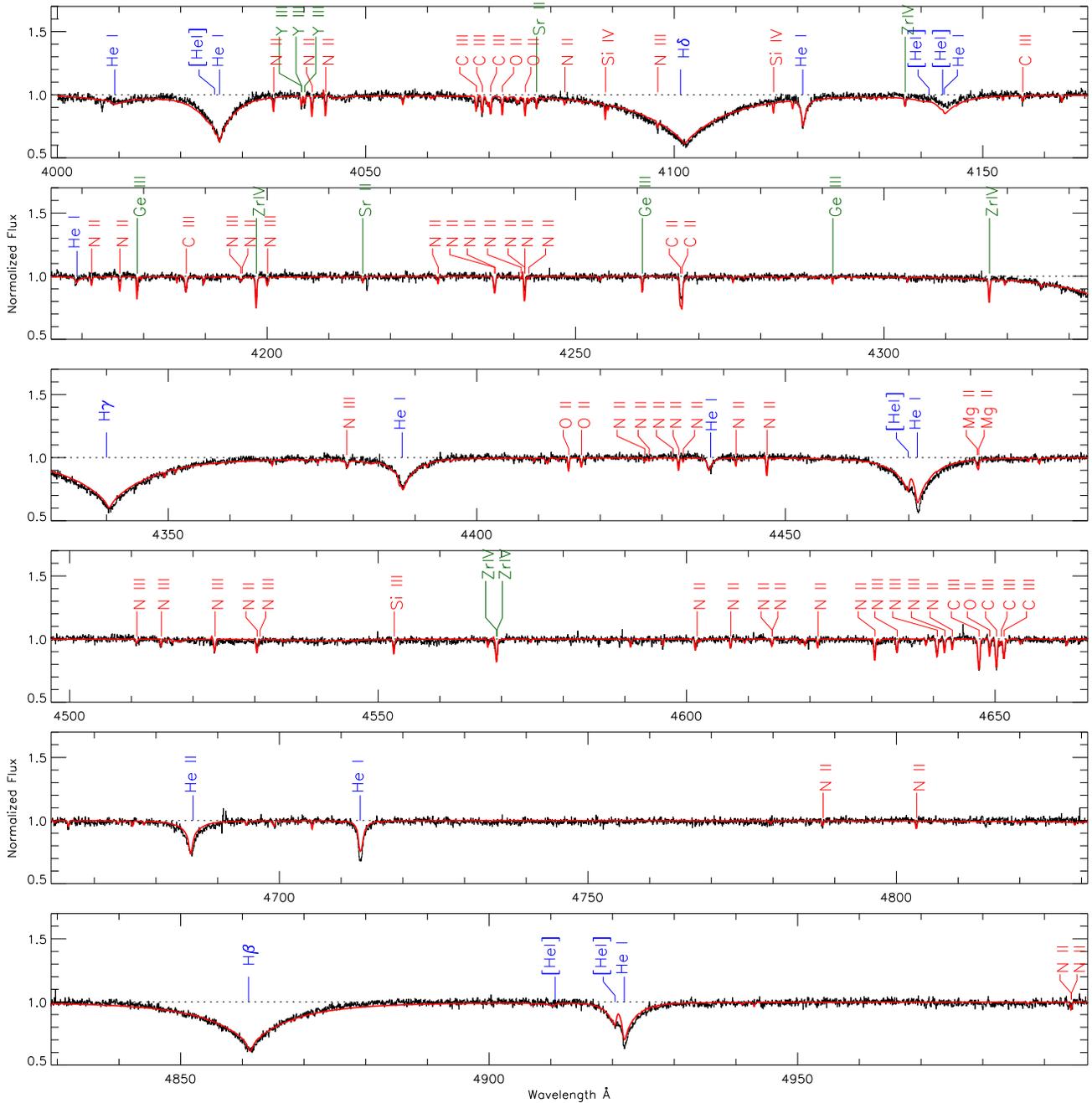}
\caption{The merged AAT/UCLES spectrum of LS\,IV$-14^{o}116$ 
  along with the best FIT model
  of $T_{\rm eff}=34\,000\,{\rm K}$ and $\log g=5.50$. Abundances are as in Table~\ref{t_abunds}.}
\label{fig_AATfit}
\end{figure*}

\begin{table}
\centering
\caption{Mean abundances $\epsilon_i$ for LS\,IV$-14^{\circ}116$.}
\label{t_abunds}
\begin{tabular}{@{}cccccc}
\hline
El. & LS\,IV$-14^{\circ}116$ & sdB$^{1-5}$ & JL\,87$^6$ & He-sdB$^7$ & Sun$^8$ \\
\hline
 H   & $11.83\pm0.07$&   12.0   & 11.6 & $<8.5-11.1$ & 12.00  \\
 He  & $11.15\pm0.05$& 7.9--11.4& 11.3 & 11.5 & 10.93  \\
 C   & $8.04\pm0.22$& 6.9--7.5 &  8.8 & 7.1--9.0  &  8.52  \\
 N   & $8.02\pm0.20$& 7.4--7.6 &  8.8 & 8.3--8.7  &  7.92  \\
 O   & $7.60\pm0.17$& 6.5--8.5 &  8.6 & 7.1--7.3  &  8.83  \\
 Ne  & $<7.6 $      & 6.5--8.5 & 8.31   &8.2--8.5 &  [8.08]  \\
 Mg  & $6.85\pm0.10$& 5.6--7.6 &  7.4 & 7.1--8.3  &  7.58  \\
 Si  & $6.32\pm0.12$& 5.5--7.7 &  7.2 & 6.8--7.4  &  7.55  \\
Ar   & $<6.5 $          & 6.0--9.0       &      &           &  [6.40]      \\
Sc   & $<5.3 $          & 5.0--7.0 &      &           &  3.17      \\
Ti   & $<6.0 $          & 5.3--9.0 &      &           &  5.02      \\
V    & $<6.5  $         & 6.0--8.5 &      &           &  4.00      \\
Cr   & $<7.0  $         & 5.5--8.0 &      &           &  5.67      \\
Fe   & $<6.8 $          & 6.5--8.1 &      &           &  7.50      \\
 Ge  & $6.28\pm0.12$& 1.5--5.5 &    &   &  3.41  \\
 Sr  & $6.96\pm0.15$&          &    &   &  2.97  \\
 Y   & $6.16\pm0.10$&          &    &   &  2.24  \\
 Zr  & $6.53\pm0.24$& 2.7--4.9 &    &   &  2.60  \\
\hline
\end{tabular}\\
\parbox{80mm}{
1. \citet{edelmann03},
2. \citet{geier08},
3. \citet{geier10},
4. \citet{otoole06}, 
5. \citet{chayer06},
6. \citet{ahmad07},
7. \citet{naslim10},
8. \citet{grevesse98} }
\end{table}

\begin{table}
\centering
\caption{Abundance errors $\delta \epsilon_i$ due to representative errors
  in $T_{\rm eff}$, $\log g$ and $v_{\rm t}$}
\label{t_err}
\begin{tabular}{@{}llll}
\hline
Element& $\delta T_{\rm eff}=1000\,\rm K$     & $\delta \log g=0.2$ & $\delta v_{\rm t}=5$ \\
\hline
 C   & +0.05   &  +0.03  & --0.11 \\
 N   & --0.03  &  +0.02  & --0.08 \\
 O   & --0.08  & --0.004 & --0.04 \\
 Mg  & --0.07  & --0.01  & --0.03 \\
 Si  & --0.13  & --0.02  & --0.04 \\
 Ge  & --0.13  & --0.02  & --0.05 \\
 Sr  & --0.24  & --0.08  & --0.01 \\
 Y   & --0.08  & --0.02  & --0.03 \\
 Zr  & +0.02   &  +0.02  & --0.11 \\
\hline
\end{tabular}
\end{table}

Individual line equivalent widths and abundances are given in Tables
\ref{t_atomic} and \ref{t_lines}\footnote{Abundances were calculated
  with improved model atmospheres, which give values different to
  those shown by \citet{naslim10}}. Abundances are given in the form
$\epsilon_i = \log n_i + c$ where $\log \Sigma_i a_i n_i = \log
\Sigma_i a_i n_{i\odot} = 12.15$ and $a_i$ are atomic weights. This
form conserves values of $\epsilon_i$ for elements whose abundances do
not change, even when the mean atomic mass of the mixture changes
substantially.

Mean abundances for each element are given in 
Table~\ref{t_abunds}. The errors given in Table~\ref{t_abunds}
are based on the standard deviation of the line abundances about 
the mean or the estimated error in the equivalent width
  measurement. Systematic shifts attributable to errors in
    $T_{\rm eff}$, $\log g$ and $v_{\rm t}$ are given in Table~\ref{t_err}.
The final best-fit spectrum using the adopted best-fit model and with
the elemental abundances from Table~\ref{t_abunds} is shown in 
Fig.~\ref{fig_AATfit}, together with identifications for all of the
absorption lines in the model. 

Table~\ref{t_abunds} also shows a representative range of abundances
measured for normal sdB stars, for the intermediate helium sdB star JL\,87,
for helium-rich sdB stars and the Sun. In common with many helium-rich sdB stars, 
LS\,IV$-14^{\circ}116$ appears to be mildly metal-poor (Mg and Si are
assumed as proxies for the mean metallicity). Relative to the Sun,
nitrogen is unremarakble, but relative to Mg and Si,
LS\,IV$-14^{\circ}116$ is arguably N-rich. 

With a helium-to-hydrogen ratio of 0.2, LS\,IV$-14^{\circ}116$
falls well between the normal sdB stars, in which helium is strongly
depleted, and the truly helium-rich sdB and sdO stars
\citep{naslim10,stroeer07}. Only JL\,87, with $n_{\rm H}/n_{\rm
  He}\approx1$ might be comparable \citep{ahmad07}. However, JL\,87 is
strongly C- and N-rich, and shows no evidence for exotic species such
as those described here. 

Heavier elements, including argon and the iron-group elements
  scandium, titanium, vanadium, and chromium are frequently seen
  to be overabundant in H-rich sdB stars with higher
  effective temperatures \citep{geier10}. LS\,IV$-14^{\circ}116$ does not show any
  detectable lines due to these elements. We estimated upper limit abundances
  assuming a minimum observable equivalent width of $5{\rm m\AA}$
  (Table~\ref{t_abunds}). Overabundances of Sc, Ti, V and Cr cannot be
  ruled out.

What is undeniable is that LS\,IV$-14^{\circ}116$ shows overabundances
of Ge, Sr, Y and Zr which are unprecedented in any hot subdwarf and,
possibly, in any other star (Sr, Y and Zr are only $\approx2$\,dex overabundant in
Przybylski's star HD101065, although heavier elements do show 4\,dex
overabundances: \citet{cowley00,shulyak10}).

\section{Discussion}

Both normal sdB stars and helium-rich subdwarfs are chemically
peculiar. The former are depleted in helium and in other elements
with $Z<20$, but enriched in all elements $Z>20$ except
iron \citep{otoole06,geier10}. The consensus explanation is that 
chemical diffusion operates in an atmosphere where gravitational
settling and radiative levitation compete to elevate ions to layers
in which their specific opacities are maximised. 
Abundance information for 
helium-rich subdwarfs is less extensive \citep{naslim10}, 
but the absence of hydrogen and the enrichment of nitrogen and (sometimes) 
carbon point to nuclear-processed helium dominating the surface
composition. A few intermediate-helium sdB stars (with $0.1 < n_{\rm
  He} < 0.9$) may be transition objects; if helium-rich sdBs are 
contracting towards the extended horizontal branch,
intermediate-helium sdB stars may represent those stars in which 
diffusion has been effective sufficiently long to deplete the surface
helium by a significant amount \citep{naslim10}. 

\subsection{Diffusion}

However, chemical peculiarities of the magnitude seen on the surface of 
LS\,IV$-14^{\circ}116$ demand a more sophisticated explanation. 
Are they produced by some diffusion process which has led to extreme
concentrations ({\it i.e.} clouds) in the line-forming region of 
the photosphere? Or are they a consequence of dredging-up heavily
nuclear processed material from the intershell region of an AGB star? 

LS\,IV$-14^{\circ}116$ would barely qualify as ``helium enriched''
were it a main-sequence star. Only because sdB stars are
generally so helium poor was it considered unusual in the first
place. So the question of whether diffusion is responsible for other
abundance anomalies is legitimate.

The elements Sr, Y, and Zr are recognized as products of s-process
neutron-capture reactions which occur in the intershell in 
asymptotic giant-branch (AGB) stars. They can appear on stellar surfaces
following either third dredge-up on the AGB, 
a very late thermal pulse ({\it e.g.} FG\,Sge, V4334\,Sge:
\cite{jeffery06,asp99}), 
or as a consequence of a white dwarf merger, 
 ({\it e.g.} V1920\,Cyg, HD124448: \cite{pandey04}).
The difficulties here are: (a) that none of these processes has previously been
indicated to produce such a large excess; 1--2 dex is more common, 
(b) the absence of a characteristic carbon and/or oxygen excess, and
(c) there is no evidence that LS\,IV$-14^{\circ}116$ is a post-AGB
star in any sense. 

Some abundances anomalies have been recognized as due to pollution
from an external source. Dust-fractionated accretion from the
interstellar medium tends to produce modest anomalies which are
correlated with condensation temperature. This is not the case here. 
Accretion from the wind of a now unseen AGB companion can also 
be ruled out by the absence of a carbon and/or oxygen excess. 

The principal advantage of the diffusion hypothesis is that it allows
for a very high overabundance of a given species in a very narrow layer of the stellar
atmosphere, but without requiring an additional source of that
material providing that there is a commensurate depletion in other layers. Since the
line-forming region of the stellar atmosphere
(optical depth $0.001\lesssim\tau\lesssim0.1$) 
contains $\approx4\pm1\times10^{-14}{\rm M_{\odot}}$ compared with an 
estimated $7\times10^{-13} {\rm M_{\odot}}$  for the total mass of the atmosphere
($\tau\lesssim10$) and $\approx 0.001 {\rm M_{\odot}}$ for the
mass of the stellar envelope, the observed excess can be easily 
understood.
It is important to note that overabundances of strontium and yttrium of up to 3 
dex are well-documented in  chemically-peculiar A (Ap) and mercury-manganese 
(HgMn) stars \citep{cowley08,cowley10,dworetsky08}, where they are 
commonly attributed to radiatively-driven diffusion producing a strongly 
stratified atmosphere.
 
\subsection{Variability}

LS\,IV$-14^{\circ}116$ is known to be a low-amplitude light variable. 
\citet{ahmad05} report two periods (1950 and 2900\,s) and suggest these 
are due to non-radial $g$-mode oscillations of high radial order, since the
periods are too long to be $p$-modes. A difficulty with this proposal 
is that $g$-modes are not predicted in sdB stars hotter than about
29\,000\,K \citep{jeffery06}; with $T_{\rm eff}=34\,000$\,K, 
LS\,IV$-14^{\circ}116$ sits in the middle of the sdB $p$-mode
instability zone \citep{charpinet01}. 
Our three consecutive AAT spectra show no evidence of variability in
line equivalent widths or radial velocity. This is probably a
consequence of the low amplitude of the variability and the exposure
times (1800\,s) being long compared with the photometric periods.

There are two possible links between photometric variability and 
unusual chemical composition. The first regards the blue edge of the
$g$-mode instability strip, which is very sensitive to the metal
abundance in the driving zone \citep{jeffery06b}. Reducing the hydrogen
abundance also tends to destabilize pulsation \citep{jeffery06a} 
but, in this case, the hydrogen abundance is not sufficiently depressed 
to make a major difference. Should the overabundances seen in Ge, Sr,
Y and Zr be reflected in other
elements and also be continuous from the atmosphere to the driving
zone at $\approx 2\times10^5$\,K then, in all probability, there would
be consequences for pulsational stability. However, as
discussed above, there is no obvious source for such an excess
throughout the stellar envelope. A strong requirement to
demonstrate that the light variability of LS\,IV$-14^{\circ}116$ is
due to pulsation is to investigate its radial-velocity
behaviour. Observations which are sufficiently sensitive to probe the
behaviour of Ge, Sr, Y, and Zr lines, as well as the H and He lines
would be indicative of differential motion within the stellar atmosphere.

The second possible link refers to observations of intermediate helium
stars on the main sequence, sometimes known as Bp(He) stars. The
class prototype is $\sigma\,$Ori\,E \citep{greenstein58}, 
a B2Vp star with a 1.1\,d variation in light, a chemically-inhomogeneous surface and a strong magnetic
field \citep{pedersen77,thomsen74,landstreet78}. 
Strong magnetic fields are frequently associated with large abundance
anomalies in early-type stars \citep{bohlender87,hunger99}. In simple terms, 
a magnetic field radically alters the opacity of a given ion. 
Specifically, by splitting energy-level degeneracies, 
Zeeman splitting increases the line opacity in a direction parallel to
the magnetic field lines. 
Consequently, radiative
levitation preferentially leads to the greatest accummulation 
of susceptible elements in regions where the magnetic field is 
orthogonal to the stellar surface, {\it i.e.} at the magnetic poles. 
Such anomalies can be further exacerbated by fractionation in the
stellar wind \citep{hunger99}. 
Strong chemical anomalies appear as ``spots'' and can lead to 
light variability \citep{townsend05}. 

We are not currently suggesting that  LS\,IV$-14^{\circ}116$ is a
classcial Bp(He) star. Its surface gravity is too high and so a
connection with the low-mass sdB stars seems more probable. However,
we are suggesting that the same magnetically-driven physics could be
responsible for the chemical anomalies. This suggestion can be tested. It requires 
that LS\,IV$-14^{\circ}116$ have a strong magnetic field, and it is
possible that the light variations could be associated with chemical
variability. High-resolution time-resolved spectroscopy would indicate
whether the surface is chemically homogeneous. Simultaneous photometry
would demostrate any link to light variability. Spectropolarimetry
would indicate a magnetic field.

\section{Conclusion}

LS\,IV$-14^{\circ}116$ is an intermediate helium-rich subdwarf B
star, with a gravity slightly lower than that of normal sdB stars, and
a surface helium abundance of about 16\% by number. Taking silicon and
magnesium as proxies for the mean metal abundance, it is slightly
metal poor (-0.8 dex) relative to the Sun. Since nitrogen is
approximately solar, and carbon and oxygen are 0.4 and 1.1 dex subsolar
respectively, the excess helium is overabundant, it is possible 
that the surface consists of a mixture of hydrogen with
CNO-processed helium. How diffusion affects C, N and O is not
clear. Neither is it currently known  whether
LS\,IV$-14^{\circ}116$ is a single star or has an undetected binary
companion. 

In conjunction with observations of other
helium-rich sdB stars \citep{naslim10}, it might be argued that 
LS\,IV$-14^{\circ}116$ was formed from either a helium white dwarf
merger, or a late helium-flash episode, in which such a mixture could 
arise.  Whether the observed helium/hydrogen ratio represents the
relative contributions of the original material is difficult to
establish; radiative levitation may already have caused a substantial
fraction of the helium to sink as the star contracts towards the
zero-age extended horizontal branch. 

What makes  LS\,IV$-14^{\circ}116$ remarkable, and distinct from any
other hot subdwarf, whether helium-rich or not, is the presence of 
lines due to germanium (Ge{\sc iii}), strontium (Sr{\sc ii}), yttrium 
(Y{\sc iii}) and zirconium (Zr{\sc iv}) in its optical spectrum. 
Measurements of these lines translate into 
overabundances of up to four orders of magnitude of these elements 
in the line-forming region of the photosphere. 

Since a nuclear origin for such overabundances seems unlikely, it is
argued that they arise due to radiatively-driven diffusion, where
elements accumulate in layers of high specific opacity and where
radiative and gravitational forces are in equilibrium. It is
conjectured that a strong magnetic field might be responsible for the extreme
overabundances observed in this case. How the unusual chemistry is linked to
the photometric variability remains to be established.

\section*{Acknowledgments}

This paper is based on observations obtained with the Anglo-Australian
Telescope. The authors are grateful to Drs Amir Ahmad and Timur
\c{S}ah\`in who made the observations and reduced the data. 
The Armagh Observatory is funded by direct grant from the Northern
Ireland Dept of Culture Arts and Leisure.

\bibliographystyle{mn2e}
\bibliography{mnemonic,LSIVS14116}

\label{lastpage}
\end{document}